# Energy efficient coherent quantum control of nitrogen vacancy (NV) spin with nanoscale magnets

Md Fahim F Chowdhury[1], Adi Jung[3], Léa La Spina[4], Ausrine Bartasyte[4,5,6], Samuel Margueron[4], Jayasimha Atulasimha[*,1,2,3]

[1]Department of Mechanical and Nuclear Engineering, Virginia Commonwealth University, Richmond, VA, USA.
[2]Department of Electrical and Computer engineering, Virginia Commonwealth University, Richmond, VA, USA.
[3]Department of Electrical Engineering and Computer Sciences, University of California, Berkeley, CA, USA.
[4]Université de Franche-Comté, CNRS, Institut FEMTO-ST, 26 rue de l`Epitaphe, 25000 Besançon, France.
[5]Université Paris-Saclay, CNRS, Centre de Nanosciences et de Nanotechnologies, 10 Bd Thomas Gobert, 91120, Palaiseau, France.
[6]Institut Universitaire de France, 103 boulevard Saint Michel, 75005 Paris, France.
*jatulasimha@vcu.edu

## Abstract

**We investigate coherent quantum control of a nitrogen vacancy (NV) center in diamond with microwave fields generated from a nanoscale magnet that is proximal to the NV center. Our results show remarkable coherent control with high contrast Rabi oscillations using nearfield microwaves from shape anisotropic nanomagnets of lateral dimensions down to 200 nm x 180 nm, driven remotely by surface acoustic wave (SAW) excitation that is at least 400 times and potentially 4 orders of magnitude more energy efficient than generating microwaves with an antenna. Furthermore, we show that varying the acoustic power driving such nanomagnets can achieve control over Rabi frequency. We also report spin-lattice relaxation time $T_1$ is $103\pm0.5$ $\mu s$, the spin-spin relaxation time $T_2$ is $1.23\pm0.29$ $\mu s$, and the Ramsey coherence time $T_2^*$ is $218\pm27$ $ns$ measured using microwave pulses generated by such nanomagnets. The use of the nanoscale magnets to implement highly localized and energy efficient coherent quantum control can replace thermally noisy microwave circuits and demonstrate a path to scalable quantum computing and sensing with NV-defects in diamond and other spin qubits.**

## Introduction

Scalable spin based and other microwave driven quantum systems require local sources of microwave pulse, which is fundamental to coherent manipulation of spin defects. This is critical for applications such as quantum sensors [1-5], many-qubit quantum processors, and quantum repeaters [6-9]. The research in quantum information science is fueled by the exponential computing capability with the increase in number of qubits in a processor unlike neuromorphic [10-11] and non-von Neumann computing [12-16]. Leveraging the principles of superposition, entanglement, and quantum interference, quantum processors can tackle complex problems beyond the reach of classical computers using algorithms like Shor's and Grover's [16-18] and more recently quantum AI [32]. Likewise, quantum sensing offers unparalleled precision in measuring magnetic fields, gravitational waves, and atomic forces, revolutionizing fields like navigation, medical imaging, and environmental monitoring [5].

Among many promising candidates for quantum bits, nitrogen vacancy center has special interest due to their exceptional stability in room temperature, high quantum coherence time, and local spin-entanglement. The coherent control of NV spins has been achieved through electron spin resonance and optical manipulation of the spin of the NV center, which often include the application of microwave frequency excitation, typically ranging 1W to as high as 10 W [19-21]. The requirement for such high excitation power not only adds complexity to incorporating the spins within a limited space but also poses the risk of significantly disturbing the neighboring spins. The NV center is increasingly favored for quantum sensing over quantum information processing due to challenges in localized, energy-efficient control. Despite high coherence times, slow single and multi-qubit gate operations hinder effective computation. One strategy for faster gates involves boosting microwave power or control-field amplitude. However, traditional methods, including generating electric currents in transmission lines suffer from dispersive behavior, leading to decoherence and crosstalk issues. While injecting more current can increase microwave power, it also induces joule heating and dephasing in nearby qubits, impeding scalability. Moreover, increasing the distance of qubits to mitigate these effects complicates entanglement protocols crucial for multi-qubit gate implementation. To localize microwave power in quantum processors, various methods have been proposed, such as using mechanical resonators, applying strain [22-24], leveraging the coupling of spin defects with nearby films via magnetoelasticity for coherent oscillation [27], and utilizing spin waves for coherent drive [26]. While such coherent magnetoelastic drive [27] using magnetic films could have different spin wave modes, nanoscale magnets are single domain states (see supplement, section 2.3) and tend to oscillate in a near single domain state when driven by voltage control, avoiding extra resonant modes that tend to be excited in thin films. This potentially enables highly coherent control of spin qubits as theoretically predicted and simulated in Ref [40]. Technologically, the nanoscale magnets allow the generation of highly confined microwaves [40] that can selectively drive a proximally located qubit but not disturb a neighboring qubit.

Challenges remain in achieving high-power excitation for low-noise measurements, despite superior spatial decay [28-29]. Magnetoelastic interaction presents a potential solution, reducing input power requirements and generating highly efficient magnetic fields [25]. However, challenges persist in coupling and high-power consumption due to elevated current density or thin film dimensions, ultimately constraining scalability and necessitating the use of nanomagnets to solve these challenges.

In this work, we report an energy-efficient method for coherent control of a single NV center using nanoscale magnets for the first time. We show that, under ambient conditions, elliptically shaped anisotropic cobalt nanomagnets, can be driven acoustically by microwave voltage application, which strongly couple with the NV center in diamond to manipulate its spin states. The Rayleigh-mode wave generated by the interdigital transducer (IDT) induces an effective magnetic field in the nanomagnet due to magnetoelastic coupling and thus inducing magnetization dynamics with very low power. The proximal location of the nanomagnet to the NV produces a large microwave field at the NV center, thus increasing the energy efficiency by at least ~400 times and potentially by 4 orders of magnitude (see supplementary information) compared to conventional antenna driven NV [19-21]. Advancing scalable quantum information processing requires densely packed spins [37–39]. Precise control over individual or ensemble spins is crucial to ensure high-fidelity multi-qubit gates which necessitates localized control fields to prevent cross-talk, which can otherwise compromise gate fidelity [37]. The control field can be localized and reduce footprint by using nanoscale magnets [40-41]. In this work, confining the large microwave field to nanoscale dimensions, we achieve reduced noise in the vicinity, thereby facilitating the scalability of spin/NV center-based quantum processors and sensors.

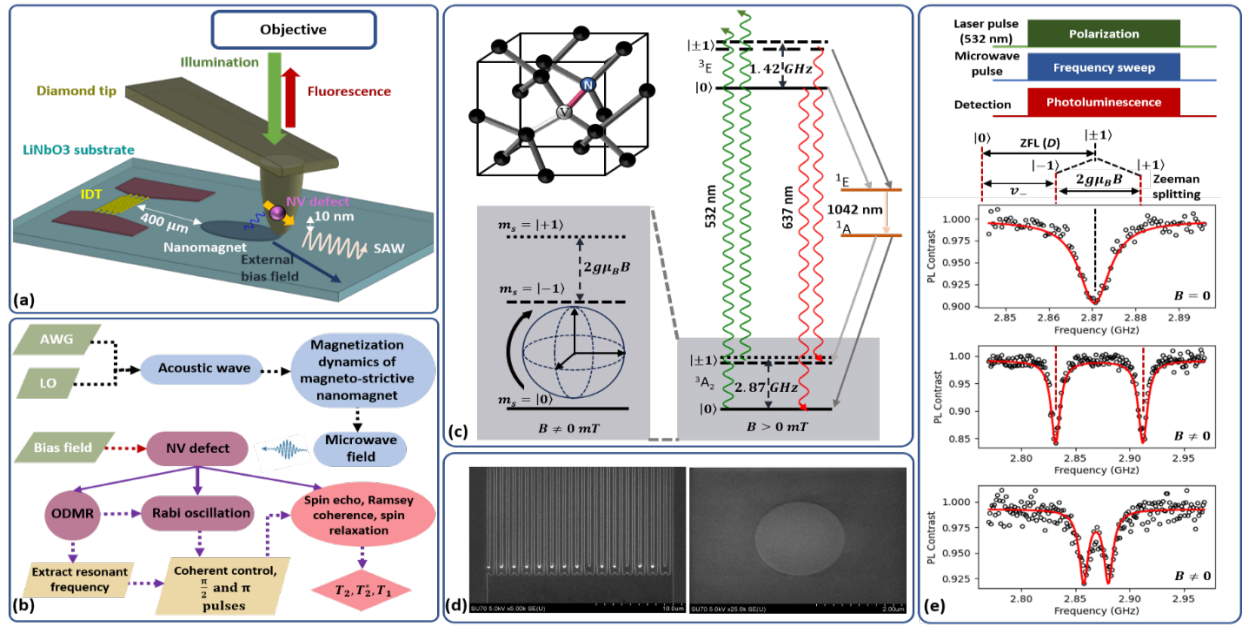

Fig. 1: Overview of coherently control NV-center with nanoscale magnets. (a) A simplified schematic diagram of the ODMR and Rabi oscillation experiment with the SAW IDT device on *LiNbO3* substrate driving a nanomagnet proximal to a diamond tip with single NV center (image not to scale). (b) Flow diagram of coherent control of NV with nanoscale magnets: experiment and measurement protocol. (c) Geometrical structure of diamond with a NV-defect. Energy level diagram of NV center with radiative and non-radiative transitions with ground, excited and metastable states. (d) Scanning electron microscopy (SEM) images of the IDT and a 2.0 $\mu m$ x 1.8 $\mu m$ nanomagnet used in the experiment. (e) Pulse sequence for continuous wave ODMR experiment. ODMR plots of the NV-center with or without the bias magnetic field used in the experiment. Abbreviation: NV-nitrogen vacancy; ODMR-optically detected magnetic resonance; SAW-surface acoustic wave; IDT-interdigital transducer; AWG-Arbitrary waveform generator.

## Results

We demonstrate coherent quantum control of NV-center with microwaves generated by a nanoscale magnet proximal to a nano-diamond (ND) tip with a NV-defect. A simplified experimental set-up can be found in Fig. 1. A device under test consists of an IDT to generate the acoustic waves, and nanomagnets with different size and anisotropy in the acoustic delay line of the IDT. The nanomagnets are placed ~400 $\mu m$ apart from the marginal IDT finger and ~200 $\mu m$ from the IDT pads where microwave voltage is applied as shown in Fig 1(a). Two wire bonds are used in each of the IDT pads to remove the parasitic inductance and ensure efficient coupling with the transmission line and SAW transducers. The microwave voltage applied to the IDT generates a Rayleigh wave in the piezoelectric substrate which propagates and induces a time varying tensile and compressive stress in the nanomagnet deposited on the piezoelectric $LiNbO_3$ substrate. Due to the magnetoelastic coupling, an effective field is induced on the magnetostrictive nanomagnet, thus converting the acoustic energy into magnetization dynamics [34-35]. The magnetization oscillation due to the Villari effect generates a large oscillating dipolar field in its proximity [36]. The NV center tip is proximal to and approximately along the easy axis of the elliptical nanomagnets such that the static global bias magnetic field is approximately along the NV axis and the microwave field generated by the oscillating magnetization has a large component perpendicular to the NV axis.

The NV center is an electronic defect in diamond with spin -1, formed by a pair of carbon vacancy and an adjacent substitutional nitrogen atom in the crystal lattice. In NV experiments, measurements of the spin states are always a projection onto the NV axis. We used a single NV center tip made from [100] cut $^{12}$C diamond, which leads to NV orientation of ~144$^o$ with respect to the normal with a nominal depth of 10 nm as shown in Fig 1. The NV center is approximately 30 nm away from the nanomagnet surface during measurement in engaged mode of the quantum scanning microscope. The negatively charged state of the defect provides a spin triplet ground level, which can be optically initialized to $m_s = |0\rangle$ state and can be coherently transitioned to the excited state $m_s = |\pm1\rangle$ by the application of resonant microwave pulses, resembling a quantum bit [33]. The all-optical readout of the NV center is performed by measuring the spin-dependent non-radiative photoluminescence (PL) emission. In the absence of a magnetic field, the degenerate $|\pm1\rangle$ levels and the zero level $|0\rangle$ are separated by $D \cong 2.87$ GHz due to magneto-crystalline anisotropy [30]. As shown in Fig. 1, we observe a drop of intensity at ~2.87 GHz in the optically detected magnetic resonance (ODMR) plot representing the zero bias field resonance of the NV-center tip with a single defect used in this experiment. Pulsed experiments with NV centers rely on detecting Zeeman shifts in spin transitions via ODMR spectroscopy [30-31]. Upon the application of a magnetic field, the $|\pm1\rangle$ states split into $m_s = |+1\rangle$ and $m_s = |-1\rangle$, which are accessible through two transition frequencies, $v_\pm$.

As shown in Fig 1(e), the Zeeman shift, defined by the effective magnetic field along the NV-axis as $2g\mu_B B_{NV}$, alters when a 5 mT bias field is applied, causing resonances at 2.830 GHz and 2.9125 GHz. As the NV center approaches the nanomagnet, the Zeeman shift reduces, suggesting an effective static field from the nanomagnets is acting opposite to the 5 mT external bias field. The static stray field from nanoscale ferromagnets varies with size, shape anisotropy, and NV center proximity to the nanomagnet. The combined effect of this static field and global magnetic field determines the Zeeman shift and the resonant frequencies for transition to $|0\rangle \leftrightarrow |-1\rangle$ or $|0\rangle \leftrightarrow |+1\rangle$ states.

**Coherent control with nanoscale magnets**

**Rabi oscillation:**

Rabi oscillation is a fundamental step towards demonstrating coherent quantum control along with other pulsed experiments, which represents an evolution of a spin under the constant drive of resonant microwave pulses. We first performed ODMR measurements to examine the NV- electron spin resonance (ESR) frequency between $|0\rangle$ and $|-1\rangle$ states in the presence of an external bias field of 5 mT applied along the NV-axis. In this continuous wave mode, the NV center is initialized by illumination of a green laser to the $|0\rangle$ state and PL is continuously measured using a confocal microscope to find the resonance frequency, $\nu_-$ for $|0\rangle \leftrightarrow |-1\rangle$ transition. This frequency is used to drive the Rabi oscillation.

The pulse sequence and Bloch vector diagram for Rabi oscillation experiment is presented in Fig. 2(a). After initializing the spin at $m_s = |0\rangle$ state, we apply an oscillating voltage at the resonant frequency characterized by the ODMR experiments to the acoustic IDT to excite the nanomagnet to generate a MW control field at the NV-center at a variable pulse duration ($\tau$) to rotate the spin to different states. The NV spin oscillates periodically between ground state, $m_s = |0\rangle$ and excited state, $m_s = |-1\rangle$ with a Rabi frequency ($T_{Rabi}$) which is dependent on the amplitude of the applied MW field generated by the nanomagnet at the NV-center spin. The oscillation is visualized by readout pulses at corresponding delays to measure the spin state $|\psi_\tau\rangle$ using a lock-in amplifier to construct the Rabi oscillation, which shows a periodic bright (high photon counts/s) and dark (low photon counts/s) PL contrast which corresponds to the $|0\rangle$ and $|-1\rangle$ states, respectively. In Fig. 2, the corresponding ODMR spectrum and the standard decaying Rabi oscillations are shown for the nanomagnets of sizes 2 μm x 1.8 μm, 1 μm x 0.9 μm, 500 nm x 470 nm, 200 nm x 180 nm. The variation of the resonant frequency of the NV- center can be attributed to the effective DC field component of the different nanomagnets' stray fields along the NV center axis. Fast Rabi frequency $T_{Rabi}=\sim 7$ MHz is achieved with a nanomagnet of size down to 200 nm and almost 28% PL contrast is measured for the 500 nm nanomagnet, which are indicative of remarkable coherent control of the NV center with a nanoscale magnet even in the presence of the room temperature background noise.

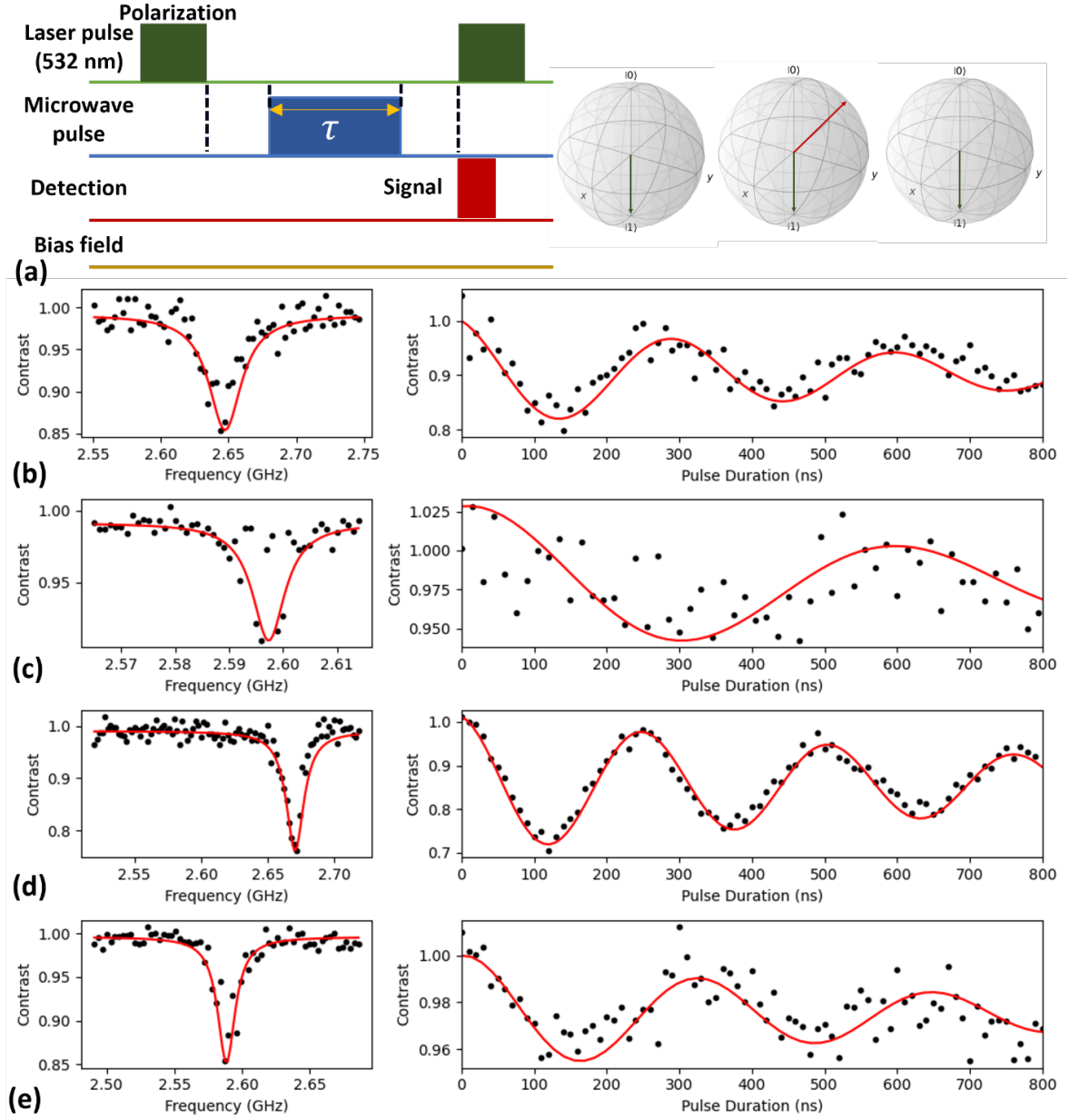

**Fig. 2: Rabi oscillation of NV center driven by the microwave pulses from different nanoscale magnets at room temperature. (a) Pulse sequence for Rabi oscillation measurement. Rabi pulse sequence is a microwave pulse resonant at the ODMR frequency with a variable length, τ between two laser pulses for polarization and readout of the spin state of the NV-center. Room temperature ODMR spectrum and corresponding Rabi oscillation on the |0⟩ to |−1⟩ transition driven by acoustically excited nanomagnets of size (b) 2.0 μm x 1.8 μm at 2.647 GHz under 5 mT bias field, (c) 1.0 μm x 0.9 μm at 2.599 GHz under 5 mT bias field, (d) 500 nm x 470 nm at 2.669 GHz under 2.4 mT bias field, and (e) 200 nm x 180 nm at 2.589 GHz under 5 mT bias field. The black scatter plots are the experimental data and red curve are the curve fitted plots.**

The linear dependency of microwave power on the Rabi frequency ($T_{Rabi}$) and single-qubit gate ($\pi/2$ and $\pi$) durations at these powers are presented in Fig. 3. The voltage amplitude of the arbitrary wave function generator (AWG) is varied to study the power dependency of Rabi frequency, while keeping the excitation frequency constant at the resonant frequency of the |0⟩ ↔ |−1⟩ state transition, i.e. the Larmor frequency. The plot in Fig. 3 shows the linear dependence of the Rabi frequency with the amplitude of microwave power. The $\pi/2$ and $\pi$-gate durations of the NV-center spin is calculated at the half and full-amplitude of the fitted Rabi oscillation curves for each power. The table and the frequency vs. AWG power factor plot in Fig. 3 demonstrate that faster gates (i.e. faster Rabi oscillation) can be achieved at higher driving pulse power, further validating our results.

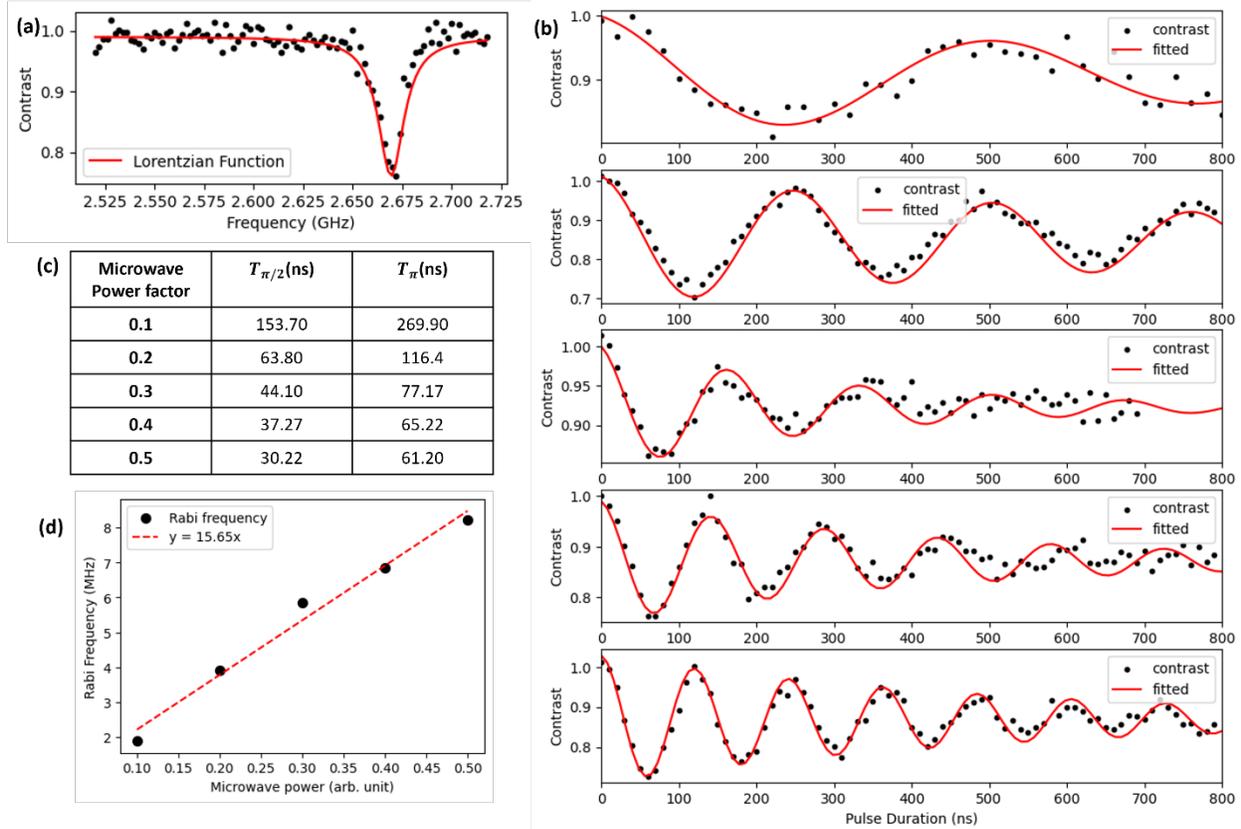

**Fig. 3: Power dependence of Rabi oscillation rate using the 500 *nm* x 470 *nm* nanomagnet. (a) Optically detected magnetic resonance plot resonant at 2.669 GHz at the proximity of 500 *nm* nanomagnet for an applied global field of 2.4 mT. (b) Rabi oscillation (black scatter plot) and fitted curve (red) at different powers driven by 500 *nm* x 470 *nm* nanomagnet at 2.669 GHz under the same bias field of 2.4 mT. Rabi rate increases from ~2 MHz to ~8 MHz with increasing microwave power. (c) Table: The $\pi/2$ and $\pi$-gate durations (*ns*) dependence to the applied microwave power from the nanomagnet power. (d) Linear dependence of Rabi frequency ($T_{Rabi}$) with microwave power applied to the IDT.**

**Spin echo measurement ($T_2$):**

The spin-spin relaxation time or transverse magnetization decay, $T_2$ can be measured by the spin echo pulse sequence experiment. We perform spin echo measurement with 2 *μm* and 200 *nm* nanomagnet devices by following the sequence $\frac{\pi}{2} - \frac{\tau}{2} - \pi - \frac{\tau}{2} - \frac{\pi}{2}$, where the $\frac{\pi}{2}$-pulse and $\pi$-pulse, are defined from the Rabi oscillation at the resonant frequency. The resonant frequency (ODMR) in cases of the 2 *μm* and 200 *nm* nanomagnets are 2.647 GHz and 2.605 GHz at 5 mT bias field, respectively due to the different bias magnetic field generated by the different sizes of nanomagnets. A $\frac{\pi}{2}$-pulse is applied on the initialized NV-center to transition to transversal plane, where the spin goes through a free evolution. A $\pi$-pulse after a delay ($\frac{\tau}{2}$) suppresses the effect of phase accumulation due to any static magnetic field. An additional $\frac{\pi}{2}$-pulse is applied following the spin-echo pulse to measure the contrast between bright state $|0\rangle$ and dark state $|-1\rangle$. The room temperature coherence time extracted from the 2 *μm* nanomagnet microwave drive is $T_{2,2\mu m}$=1.23±0.29 *μs* at 5 mT external bias field and from the 200 nm nanomagnet microwave drive is $T_{2,200nm}$=1.13±0.17 *μs* at 5 mT field. This ~10% difference in the $T_2$ time is likely due to the dephasing caused by interaction with the nanomagnet whose magnetization fluctuates at room temperature. The pulse sequence and the experimental data of the room temperature NV-spin dynamics are presented in Fig. 4. The experimental data is fitted with a decaying exponential function (red curves). The measured $T_2$ times are comparable in both cases and independent of the resonant microwave pulse and the static magnetic field from the nanomagnets.

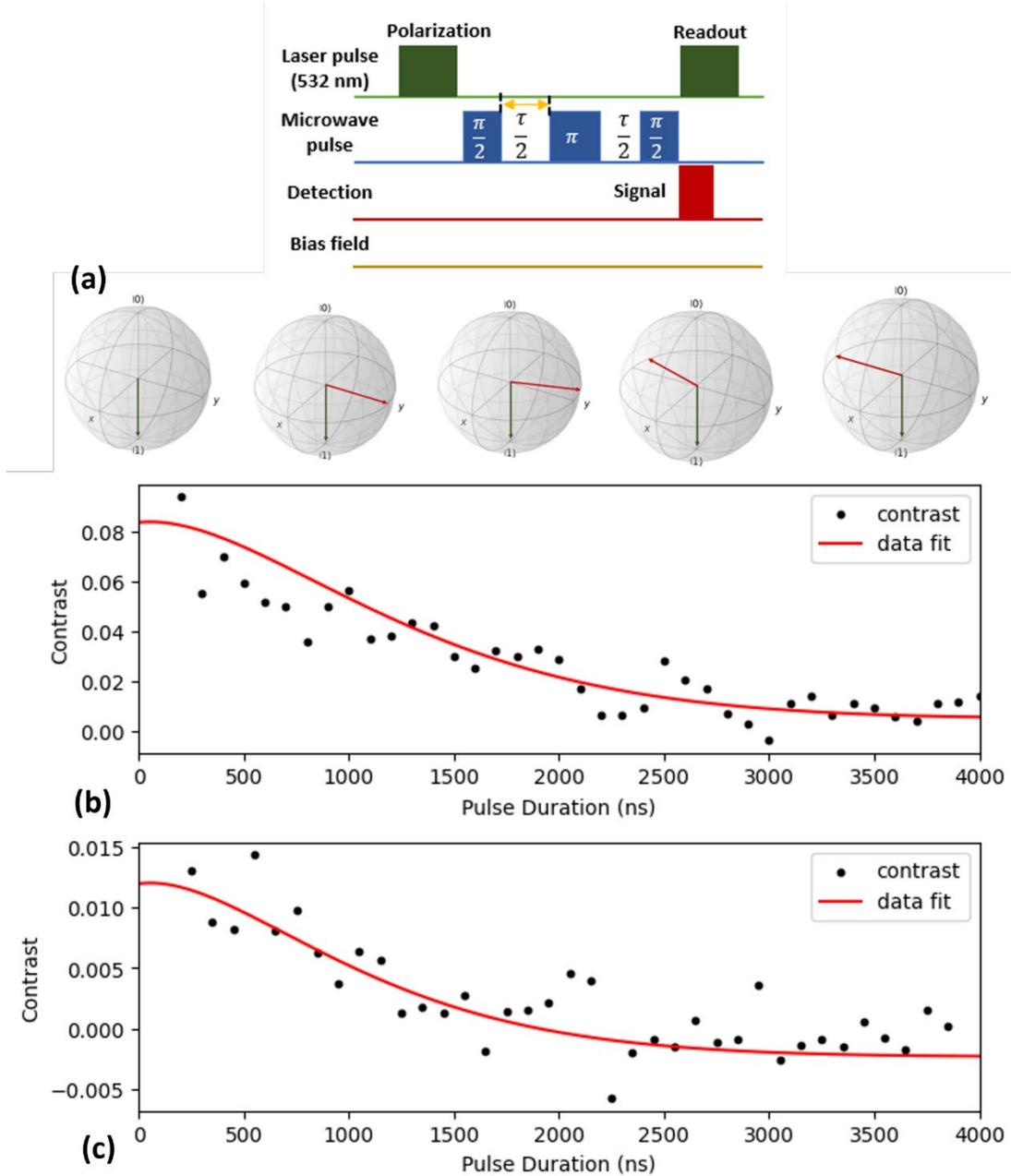

**Fig. 4: Spin-spin relaxation ($T_2$ coherence time) dynamics of NV-center spin in diamond at room temperature. (a)** Pulse sequence $(\frac{\pi}{2} - \frac{\tau}{2} - \pi - \frac{\tau}{2} - \frac{\pi}{2})$ for spin echo measurement. The NV-spin is transitioned to transversal plane with a $\frac{\pi}{2}$-pulse to go through a free evolution for $\frac{\tau}{2}$-time delay followed by a $\pi$-pulse to suppress the effect of dephasing due to static magnetic field in the system. The evolution of the NV-center quantum state is characterized by applying another $\frac{\pi}{2}$-pulse at $\frac{\tau}{2}$-delay and measuring the PL intensity. **(b)** Optically detected spin echo when driven with a 2 μm x 1.8 μm device at 2.647 GHz resonant frequency at 5 mT and **(c)** with a 200 nm x 180 nm device at 2.605 GHz resonant frequency at 5 mT.

**Ramsey interference ($T_2$*):**

To obtain the dephasing time ($T_2$*) of the NV center with our nanomagnet, we perform the Ramsey interferometry. Fig. 5 shows the pulse protocol for Ramsey experiment and the measurement result of Ramsey oscillation of the single NV-spin driven by a 2 μm x 1.8 μm nanomagnet generated microwave pulse at the ODMR frequency. Similar to other pulsed experiments, we initialized the NV-spin into the $|0\rangle$ state with a green laser pulse. The first $\frac{\pi}{2}$-pulse drives the spin to a superposition state: $|\psi\rangle = \frac{1}{\sqrt{2}}(|0\rangle + |1\rangle)$, which goes through a free spin evolution on the equator of the Bloch sphere due to the presence of the bias field. The Ramsey plot is obtained by probing the evolution of the quantum state by applying the second $\frac{\pi}{2}$-pulse with a different delay time, $\tau$. The Ramsey coherence time $T_2$* is 218±27 $ns$ and signal decays faster after one oscillation. The higher decoherence rate of the NV center can be attributed to thermal noise in the nanomagnet sample and/or in the NV spin at room temperature. The decoherence time can also be dependent on the different bias field conditions in the experiment.

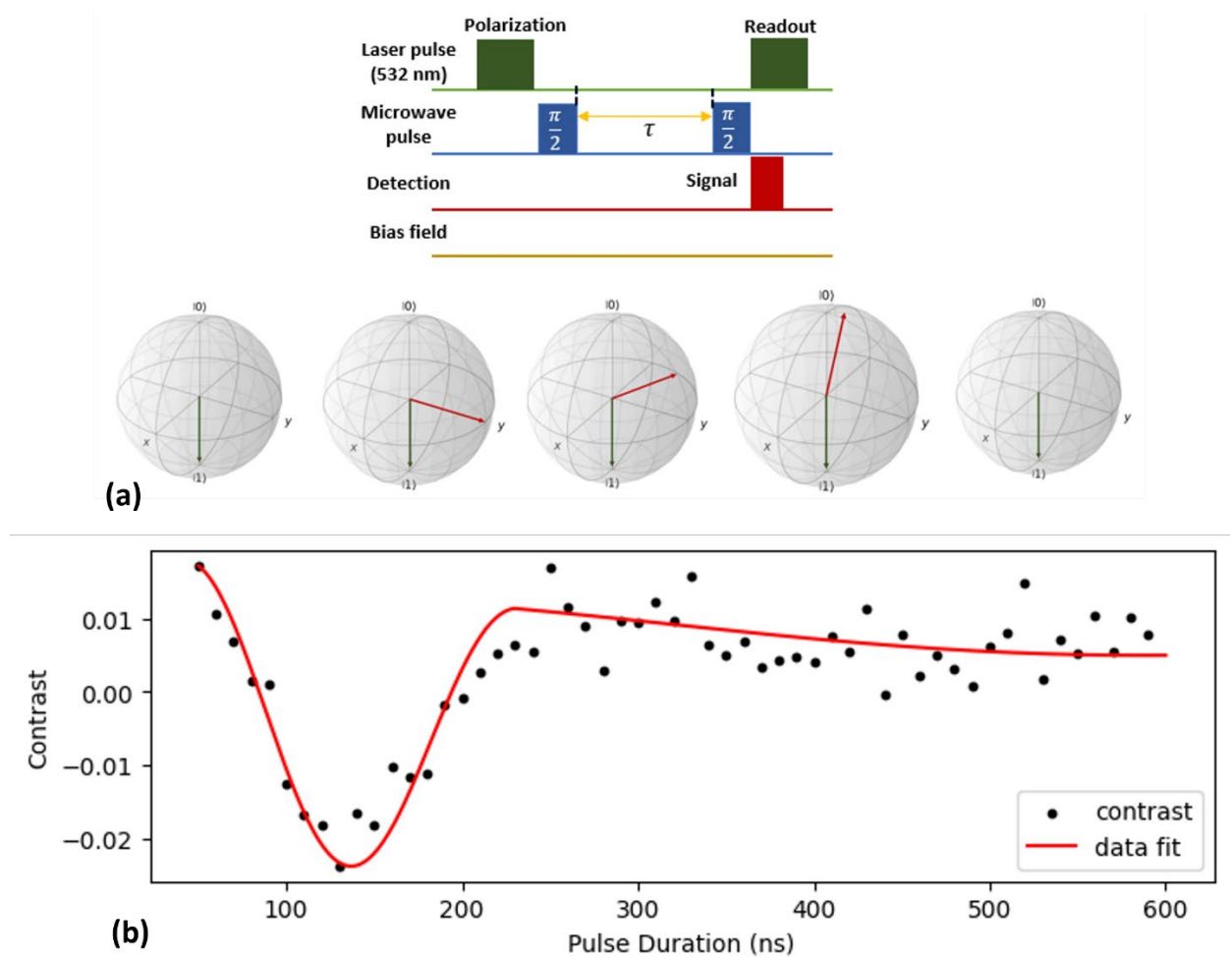

**(a)**

**(b)**

**Fig. 5: Ramsey interference experiment to measure $T_2^*$ coherence time of the NV center with nanomagnet induced microwave field at room temperature. (a) Pulse sequence of Ramsey protocol $\frac{\pi}{2} - \tau - \frac{\pi}{2}$. The NV-spin is polarized to ground state ($|0\rangle$) by applying a green laser. Two $\frac{\pi}{2}$-pulses are applied with a delay, $\tau$ between the pulses to prepare the spin in a superposition state to go through a free evolution time and then to measure the spin state. (b) Ramsey interference plot driven by a 2.0 μm x 1.8 μm nanomagnet with resonance at 2.692 GHz at 5.0 mT external bias field.**

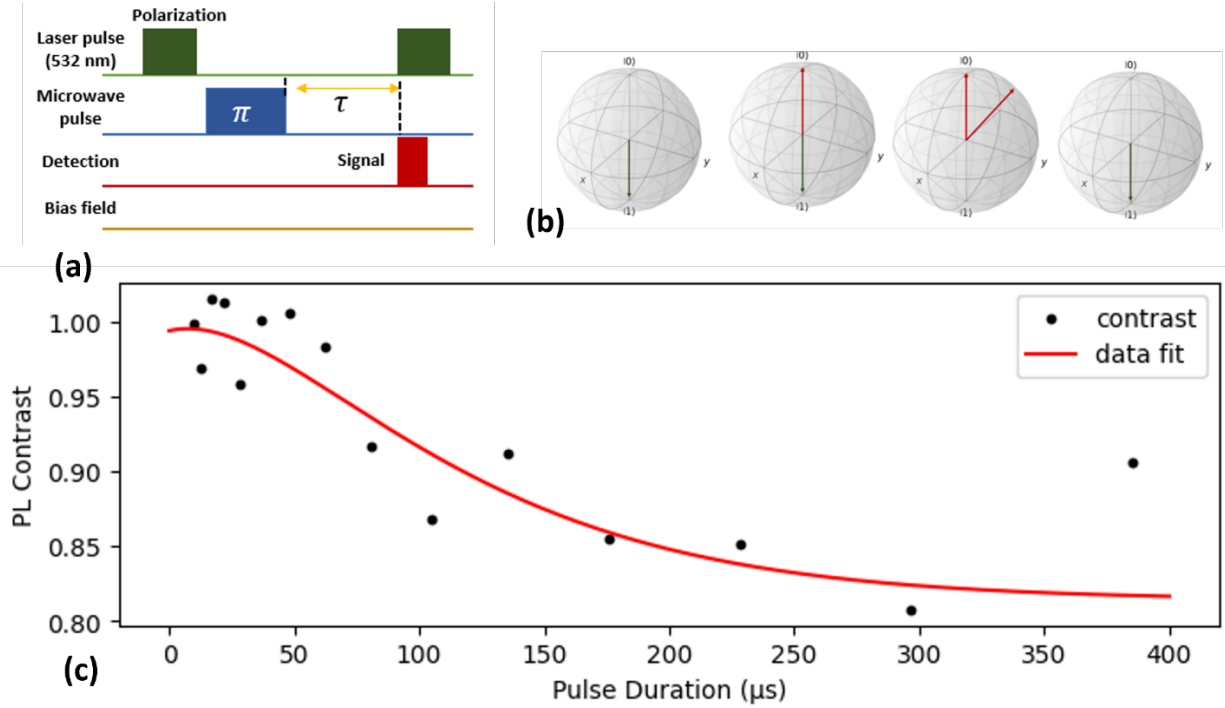

**(a)**

**(b)**

**(c)**

**Fig. 6: Spin relaxation of NV center when driven by 2.0 $\mu m$ x 1.8 $\mu m$ nanomagnet at 2.692 GHz under 5 mT bias field. (a) Pulse sequence for a spin-lattice relaxation time ($T_1$) measurement. The green indicates the laser is on and the NV center is initialized as $m_s = |0\rangle$ of the ground energy state. The NV center is allowed to relax between the $\pi$ −pulse and measurement pulse with a variable delay time $\tau$. (b) Bloch sphere representation of the spin state of the NV center. (c) Spin dependent photoluminescence contrast (black scatter plot) and fitted curve as a function of the relaxation time duration.**

### Spin-lattice relaxation ($T_1$):

The spin-lattice relaxation time ($T_1$) is measured utilizing the $\pi$-pulse, obtained from the Rabi oscillation driven by the 2 $\mu m$ x 1.8 $\mu m$ nanomagnet excited at the resonant frequency of 2.692 GHz at 5 mT global field. Fig. 6 illustrates the pulse sequence (a) and the corresponding Bloch sphere representation (b) used to characterize spin relaxation dynamics. The spin-relaxation time or the longitudinal relaxation time ($T_1$) defines the upper limit for spin coherence and is a measure of rate of reaching the spin ground state in the direction of the effective magnetic field in the system. The NV-spin is initialized at the ground state $|0\rangle$ state by applying a laser pulse and a $\pi$-pulse to transition to excited state $|-1\rangle$ subsequently. The spin state at $|-1\rangle$ undergoes free evolution on the Bloch sphere after the pulse and measured with varying free evolution time $\tau$. The free evolution is plotted in Fig. 6 and from the fitted curve the relaxation time $T_1$ is measured as 103±0.5 $\mu s$.

### Discussion

By using a magnetostrictive nanomagnet whose magnetization is driven by SAW through magnetoelastic coupling, we have demonstrated coherent quantum control of a proximally located NV-center defect in diamond. We have performed different pulse protocols such as ODMR, Rabi oscillation, spin-spin relaxation ($T_2$), Ramsey sequence ($T_2^*$), spin-lattice relaxation ($T_1$) with the microwave drives generated by nanoscale magnets at room temperature under ambient conditions. Such microwaves produced by nanomagnets are at least ~400 times and potentially 4 orders of magnitude more energy efficient than those generated by an electromagnetic (EM) antenna to obtain the same Rabi oscillation frequency, even with an unoptimized acoustically coupled nanomagnet system. The wire antenna is positioned about 50 μm away from the NV center, whereas the IDT fingers for generating acoustic waves are situated nearly 400 μm from the nanomagnet. We note that the nanomagnet itself is proximal to the NV center (the NV is within 50 nm height from the nanomagnet surface). Had the wire antenna been positioned 400 μm away from the NV, the power needed would have been 4 orders of magnitude higher.

The bottleneck of the NV centers becoming a viable candidate for quantum computing and sensing is the limit of scalability due to the dispersive control field and high current density required to manipulate the spin to implement fast quantum gates and other sensing protocols. Our work demonstrates that the use of nanomagnets for coherent control of proximally located NV and spin qubits has the potential to solve these problems and lead to scalable quantum computing and quantum sensing with NV centers in diamond by potentially replacing inefficient antennas currently used for such quantum control.

Thus, this works brings together the state-of-the-art energy efficiency in nanomagnetism/spintronics towards enabling advances in NV center-based quantum computing and sensing and is the first demonstration of using nanomagnets to coherently control a NV-defect. It can stimulate further research on use of nanomagnets with different magnetic materials for higher magnetoelastic coupling, magnetoelectric coefficient/ voltage control of magnetic anisotropy (VCMA) coefficients, higher magnetization and lower damping, which can be scaled to tens of nanometer lateral dimension or driven with local strain or VCMA instead of SAW to enable further increase in energy efficiency and downscaling.

## Methods

**Device description:** The interdigitated transducers (IDTs) and nanoscale magnets are fabricated on a (YX$l$)/128° $LiNbO_3$ substrate. The IDTs are made of 80 nm thick Al film and designed to have a nominal resonant frequency of 2.6 GHz. The nanomagnets is patterned with electron beam lithography on a bilayer polymethyl methacrylate (PMMA) followed by deposition of a 4 nm Ti adhesion layer and a 14 nm of Co by means of e-beam evaporation technique at pressure $2.2x10^{-7}$ Torr. The free standing nanomagnets are realized after lift-off.

**Curve-fitting of the pulse experiment data:** For the ODMR plots shown in the manuscript, we used Lorentzian function for data fitting. Lorentzian function, $L$, is defined by the equation below and can be generally used for background pre-processing and for spectral intensity curve fitting:

$$L(x) = \frac{1}{\pi} \frac{\gamma}{(x-x_0)^2 + \gamma^2} \tag{1}$$

Here, $\gamma$ and $x_0$ are constants defining the width and centre of the curve, respectively. The general Lorentzian function is rescaled to the corresponding frequency data and photoluminescence intensity data to fit the ODMR plot of the NV tip.

In the presence of magnetic field, the NV-center experiences the Zeeman splitting and can be observed in the ODMR spectrum with resonances in two different frequencies. This curve is fitted by the addition of two Lorentzian functions with $x, x_0\ and\ \gamma$-values of the corresponding resonances:

$$L_1(x^1) + L_2(x^2) = \frac{1}{\pi} \left[ \frac{\gamma_1}{(x-x_{01})^2 + \gamma_1^2} + \frac{\gamma_2}{(x-x_{02})^2 + \gamma_2^2} \right] \tag{2}$$

Here, $x_{01}$ and $\gamma_1$ are centre and width corresponding to the first resonant frequency, $\upsilon_-$ and $x_{02}$ and $\gamma_2$ are centre and width corresponding to the second resonant frequency, $\upsilon_+$.

The Rabi oscillation data is fitted by equation below:

$$f(\tau) = Ae^{-\tau/T_a} \cos(2\pi f\tau + \varphi) + c \tag{3}$$

Here, $A$ is the amplitude, $e^{-\tau/T_a}$ defines the exponential decay of the envelope of the oscillation where $\tau$ is the pulse duration, $(1/T_a)$ is the decay rate, $f$ is the frequency, $\varphi$ is the phase shift and $c$ is the offset.

**Acknowledgement:** M.F.F.C and J.A. were supported by the US National Science Foundation (NSF) ExpandQISE grant # 2231356. This work was supported by the French RENATECH network and its FEMTO-ST technological facility and C2N micro nanotechnologies platform, the French national ANR projects MAXSAW ANR-20-CE24-0025, and the graduate school EUR EIPHI contract ANR-17-EURE-0002.

After J. A conceived this idea, he had a preliminary discussion with Prof. Sayeef Salahuddin who was his host during his sabbatical at Electrical and Computer Engineering in UC Berkeley. J. A. would like to thank Prof. Salahuddin whose group had worked on magnetoelastic coupling to NVs and particularly discussion and presentation of results by his student Adi Jung on coherent NV control with SAW driven magnetic thin films about two months before their publication in his PhD thesis (https://escholarship.org/uc/item/1qg9v8d2) in December 2023. Furthermore, Prof. Salahuddin's group provided J.A. and M.F.F.C. unused Lithium Niobate samples with IDTs that were made by L.L.S., A. B. and S.M.

**Author contributions:** J.A. conceived the idea. This was further developed and experiments were planned by J.A., M.F.F.C. and A.J. The nanomagnets and contacts were fabricated by M.F.F.C. with inputs from A.J. on *LiNbO₃* substrates with IDTs that were fabricated by L. L. S. and A. B. with inputs from S.M. The NV measurements at QZabre were performed by M.F.F.C. with constant discussion with J.A. M.F.F.C. and J.A. wrote the manuscript and A.J., A. B. and S.M. commented on the manuscript.

## References

1. Maze, J. et al. Nanoscale magnetic sensing with an individual electronic spin in diamond. *Nature,* **455**, 644–647 (2008).
2. Balasubramanian, G. et al. Nanoscale imaging magnetometry with diamond spins under ambient conditions. *Nature*, **455**, 648–651 (2008).
3. Taylor, J. et al. High-sensitivity diamond magnetometer with nanoscale resolution. *Nature Phys.,* **4**, 810 816 (2008).
4. Aslam, N. et al. Nanoscale nuclear magnetic resonance with chemical resolution. *Science*, **357**, 67-71 (2017).
5. Degen, C. L., Reinhard, F. & Cappellaro, P. Quantum sensing. Rev. Mod. Phys., 89, 035002 (2017).
6. Kimble, H. J. The quantum internet. *Nature,* **453**, 1023–1030 (2008).
7. Munro, W. J., Azuma, K., Tamaki, K. & Nemoto, K. Inside quantum repeaters. *IEEE J. Sel. Top. Quantum Electron.,* **21**, 78–90 (2015).
8. Simon, C. Towards a global quantum network. *Nat. Photon.,* **11**, 678–680 (2017).
9. Pompili, M., et al. Realisation of a multimode quantum network of remote solid-state qubits. *Science*, **372**, 259-264 (2021).
10. Monroe, D. Neuromorphic computing gets ready for the (really) big time. Commun. ACM 57, 13 (2014).
11. del Valle, J. et al. Subthreshold firing in Mott nanodevices. Nature 569, 388–392 (2019).
12. Von Neumann, J. First draft of a report on the EDVAC. IEEE Ann. Hist. Comput. 15, 27–75 (1993).
13. V. K. Sangwan and M. C. Hersam, Neuromorphic nanoelectronic materials, Nature nanotechnology **15**, 517 (2020).
14. A. Sebastian, M. Le Gallo, R. Khaddam-Aljameh, and E. Eleftheriou, Memory devices and applications for in-memory computing, Nature nanotechnology **15**, 529 (2020).
15. C. Liu, H. Chen, S. Wang, Q. Liu, Y.-G. Jiang, D. W. Zhang, M. Liu, and P. Zhou, Two-dimensional materials for next-generation computing technologies, Nature Nanotechnology **15**, 545 (2020).
16. A. Montanaro, Quantum algorithms: an overview, npj Quantum Information **2**, 1 (2016).
17. R. Cleve, A. Ekert, C. Macchiavello, and M. Mosca, Quantum algorithms revisited, Proceedings of the Royal Society of London. Series A: Mathematical, Physical and Engineering Sciences **454**, 339 (1998).
18. K. Bharti, A. Cervera-Lierta, T. H. Kyaw, T. Haug, S. Alperin-Lea, A. Anand, M. Degroote, H. Heimonen, J. S. Kottmann, T. Menke, *et al.*, Noisy intermediatescale quantum algorithms, Reviews of Modern Physics **94**, 015004 (2022).
19. C. C. A. Teale, "Magnetometry with ensembles of nitrogen vacancy centers in bulk diamond," thesis, Massachusetts Institute of Technology (2015).
20. H. Clevenson, M. E. Trusheim, C. Teale, T. Schröder, D. Braje, D. Englund, Broadband magnetometry and temperature sensing with a light-trapping diamond waveguide. *Nat. Phys.* **11**, 393–397 (2015).
21. A. Brenneis, L. Gaudreau, M. Seifert, H. Karl, M. S. Brandt, H. Huebl, J. A. Garrido, F. H. Koppens, A. W. Holleitner, Ultrafast electronic readout of diamond nitrogen-vacancy centres coupled to graphene. *Nat. Nanotechnol.* **10**, 135–139 (2015).
22. Barfuss, A., Teissier, J., Neu, E., Nunnenkamp, A. & Maletinsky, P. Strong mechanical driving of a single electron spin. Nat. Phys. 11, 820–824 (2015).
23. MacQuarrie, E. R., Gosavi, T. A., Jungwirth, N. R., Bhave, S. A. & Fuchs, G. D. Mechanical spin control of nitrogen-vacancy centers in diamond. Phys. Rev. Lett. 111, 227602 (2013).
24. Ovartchaiyapong, P., Lee, K. W., Myers, B. A. & Jayich, A. C. B. Dynamic strain mediated coupling of a single diamond spin to a mechanical resonator. Nat. Commun. 5, 4429 (2014).

25. Labanowski, D. et al. Voltage-driven, local, and efficient excitation of nitrogen vacancy centers in diamond. Sci. Adv. 4, eaat6574 (2018).

26. Wang, X., Xiao, Y., Liu, C., Lee-Wong, E., McLaughlin, N.J., Wang, H., Wu, M., Wang, H., Fullerton, E.E. and Du, C.R., 2020. Electrical control of coherent spin rotation of a single-spin qubit. npj Quantum Information, 6(1), p.78.

27. A. Jung, Magnetoelasticity for Integration of Quantum Defects, Ph.D. thesis, EECS Department, University of California, Berkeley (2024).

28. C. Du, T. van der Sar, T. X. Zhou, P. Upadhyaya, F. Casola, H. Zhang, M. C. Onbasli, C. A. Ross, R. L. Walsworth, Y. Tserkovnyak, Control and local measurement of the spin chemical potential in a magnetic insulator. *Science* **357**, 195–198 (2017).

29. T. van der Sar, F. Casola, R. Walsworth, A. Yacoby, Nanometre-scale probing of spin waves using single electron spins. *Nat. Commun.* **6**, 7886 (2015).

30. R. Schirhagl, K. Chang, M. Loretz, and C. L. Degen, "Nitrogen-vacancy centers in diamond: Nanoscale sensors for physics and biology," Annu. Rev. Phys. Chem. 65, 83 (2014).

31. D. Budker and M. Romalis, "Optical magnetometry," Nat. Phys. 3, 227 (2007).

32. Schuld, M., Sinayskiy, I., & Petruccione, F. (2015). An introduction to quantum machine learning. *Contemporary Physics*, *56*(2), 172-185.

33. Togan, E., Chu, Y., Trifonov, A.S., Jiang, L., Maze, J., Childress, L., Dutt, M.G., Sørensen, A.S., Hemmer, P.R., Zibrov, A.S. and Lukin, M.D., 2010. Quantum entanglement between an optical photon and a solid-state spin qubit. *Nature*, *466*(7307), pp.730-734.

34. Sampath, V., D'Souza, N., Bhattacharya, D., Atkinson, G.M., Bandyopadhyay, S. and Atulasimha, J., 2016. Acoustic-wave-induced magnetization switching of magnetostrictive nanomagnets from single-domain to nonvolatile vortex states. Nano Letters, 16(9), pp.5681-5687.

35. Bandyopadhyay, S., Atulasimha, J. and Barman, A., 2021. Magnetic straintronics: Manipulating the magnetization of magnetostrictive nanomagnets with strain for energy-efficient applications. *Applied Physics Reviews*, *8*(4).

36. Niknam, M., Chowdhury, M.F.F., Rajib, M.M., Misba, W.A., Schwartz, R.N., Wang, K.L., Atulasimha, J. and Bouchard, L.S., 2022. Quantum control of spin qubits using nanomagnets. *Communications Physics*, *5*(1), p.284.

37. F. Jelezko, T. Gaebel, I. Popa, M. Domhan, A. Gruber, and J. Wrachtrup, Observation of coherent oscillation of a single nuclear spin and realization of a two-qubit conditional quantum gate, Physical Review Letters 93, 130501 (2004).

38. J. J. Pla, K. Y. Tan, J. P. Dehollain, W. H. Lim, J. J. Morton, F. A. Zwanenburg, D. N. Jamieson, A. S. Dzurak, and A. Morello, High-fidelity readout and control of a nuclear spin qubit in silicon, Nature 496, 334 (2013).

39. P. Willke, Y. Bae, K. Yang, J. L. Lado, A. Ferr´on, T. Choi, A. Ardavan, J. Fern´andez-Rossier, A. J. Heinrich, and C. P. Lutz, Hyperfine interaction of individual atoms on a surface, Science 362, 336 (2018).

40. M. Niknam, M. F. F. Chowdhury, M. M. Rajib, W. A. Misba, R. N. Schwartz, K. L. Wang, J. Atulasimha, and L.-S. Bouchard, Quantum control of spin qubits using nanomagnets, Communications Physics 5, 284 (2022).

41. Chowdhury, M.F.F., Niknam, M., Rajib, M.M., Bouchard, L.S. and Atulasimha, J., 2023. Proximal quantum control of spin and spin ensemble with highly localized control field from skyrmions. arXiv preprint arXiv:2401.00573.

# Supplementary information for

# Energy efficient coherent quantum control of nitrogen vacancy (NV) spin with nanoscale magnets

Md Fahim F Chowdhury[1], Adi Jung[3], Léa La Spina[4], Ausrine Bartasyte[4,5,6], Samuel Margueron[4],

Jayasimha Atulasimha[*,1,2,3]

[1]Department of Mechanical and Nuclear Engineering, Virginia Commonwealth University, Richmond, VA, USA.

[2]Department of Electrical and Computer engineering, Virginia Commonwealth University, Richmond, VA, USA.

[3]Department of Electrical Engineering and Computer Sciences, University of California, Berkeley, CA, USA.

[4]Université de Franche-Comté, CNRS, Institut FEMTO-ST, 26 rue de l'Epitaphe, 25000 Besançon, France.

[5]Université Paris-Saclay, CNRS, Centre de Nanosciences et de Nanotechnologies, 10 Bd Thomas Gobert, 91120, Palaiseau, France.

6Institut Universitaire de France, 103 boulevard Saint Michel, 75005 Paris, France.

*jatulasimha@vcu.edu

### Table of contents for supplementary information:

## 1. Design and fabrication of interdigitated transducer and nanoscale-magnets:

**Device description:**

The interdigital transducers (IDTs) and nanoscale magnets are fabricated on a 128 Y-cut, X-propagating lithium niobate substrate. The nanoscale magnets are fabricated around 400 μm from the IDT fingers along the aperture of the IDT. A schematic diagram of the fabrication steps for the nanomagnets are given in supplementary figure S1. The LiNbO$_3$ substrate is cleaned to remove any organic residue and spin coated with 2 layers of polymethyl methacrylate (PMMA) of different densities. The first layer is coated with ~180 nm of PMMA 495 A4 and baked for 60 seconds at 180 degree Celsius. The second layer is coated with ~ 200 nm of PMMA 950 A4 and baked for 60 seconds at 180 degree Celsius. The PMMA resist is patterned in a Raith Voyager electron beam lithography with a column of 50 kV. The patterned PMMA is developed by spray and rinse for 45 seconds with a mixture of MIBK and IPA of ratio 1:3, followed by another 45 seconds of IPA rinse and spray. The sample is post-develop baked with a hot plate at 100 degree Celsius for 60 seconds. A 4 nm Titanium (Ti) is deposited as an adhesion layer followed by a 14 nm Cobalt (Co) magnetic layer in electron-beam evaporation chamber without breaking the vacuum at $2.2 \times 10^{-7}$ Torr pressure. Finally, the PMMA resist lift-off is done with a 70-degree Celsius Remover PG solution on a hot plate with the sample immersed and a magnet bar rotating with 120 round per minute (rpm). The scanning electron microscopy images of the nanomagnets are shown in supplementary figure S1(c-e). The IDTs are made of 80 nm thick Aluminum and designed to have a resonant frequency of 2.6 GHz. Two wires are bonded to each IDT pad to ensure the proper electrical contact between the pads and microwave port transmission line. The scanning electron microscopy images of the IDT fingers are shown in supplementary figure S1(f-h).

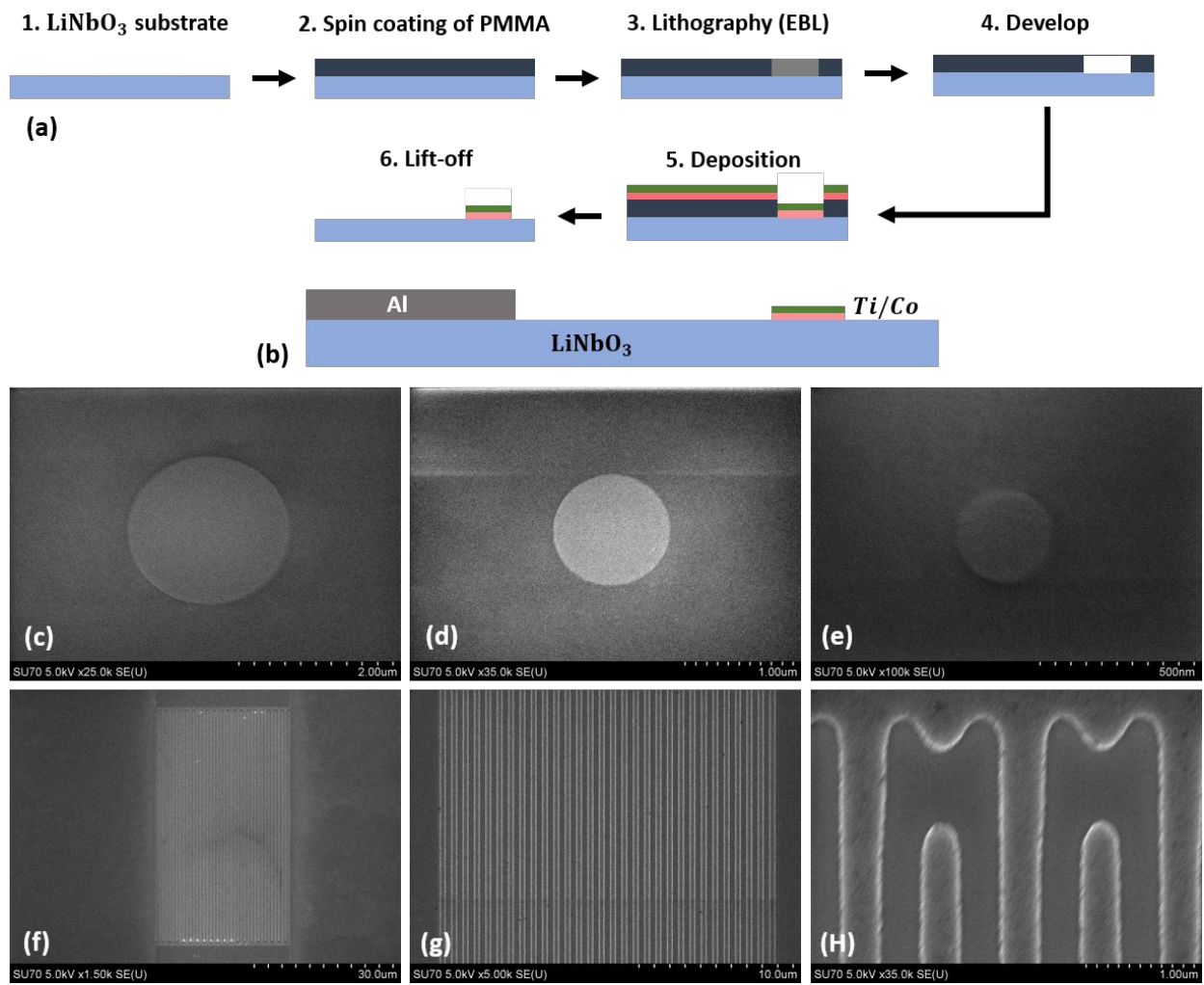

**Figure S1:** **(a) A schematic of the cross sections corresponding to fabrication steps of the nanomagnets on lithium niobite substrate. Two PMMA layers with different thickness are spin coated and post-baked. The resist layers are patterned with e-beam lithography. The resist is developed with a mixture of MIBK and IPA of 1:3 ratio. The deposition of Ti and Co is done with e-beam evaporation with thickness 4 nm and 14 nm respectively. The lift-off is done with Remover PG. (image not in scale) (b) A schematic of the cross section of the final device with IDT and nanomagnet. (image not in scale). (c) SEM image of a 2.0 µm x 1.8 µm nanomagnet. (d) SEM image of a 1.0 µm x 0.9 µm nanomagnet. (e) SEM image of a 200 nm x 180 nm nanomagnet.**

## 2. IDT and nanomagnet characterization:

### 2.1 IDT scattering parameters:

We measure the reflection coefficient of the fabricated IDT that is used for the coherent control of a spin defect in diamond with nanoscale magnets experiment. The return loss gives the information about the operating range of the frequency of the IDT. The return loss is defined by $RL = -S_{11} = -10\log\left(\left|\frac{P_{ref}}{P_{fwd}}\right|\right) = -10\log\left(\left|\frac{V_{ref}}{V_{fwd}}\right|\right) = -20\log(|\Gamma|)$, where $\Gamma$ is reflection coefficient. The measured data shown in supplementary figure S2 shows a broad bandwidth of transmission of SAW wave including the resonant frequencies of the NV-center under a bias field during the pulsed ODMR measurements. In the plot, 0 dB indicates that the entire signal is reflected, while -10 dB signifies that approximately 90% of the signal is transmitted through the channel.

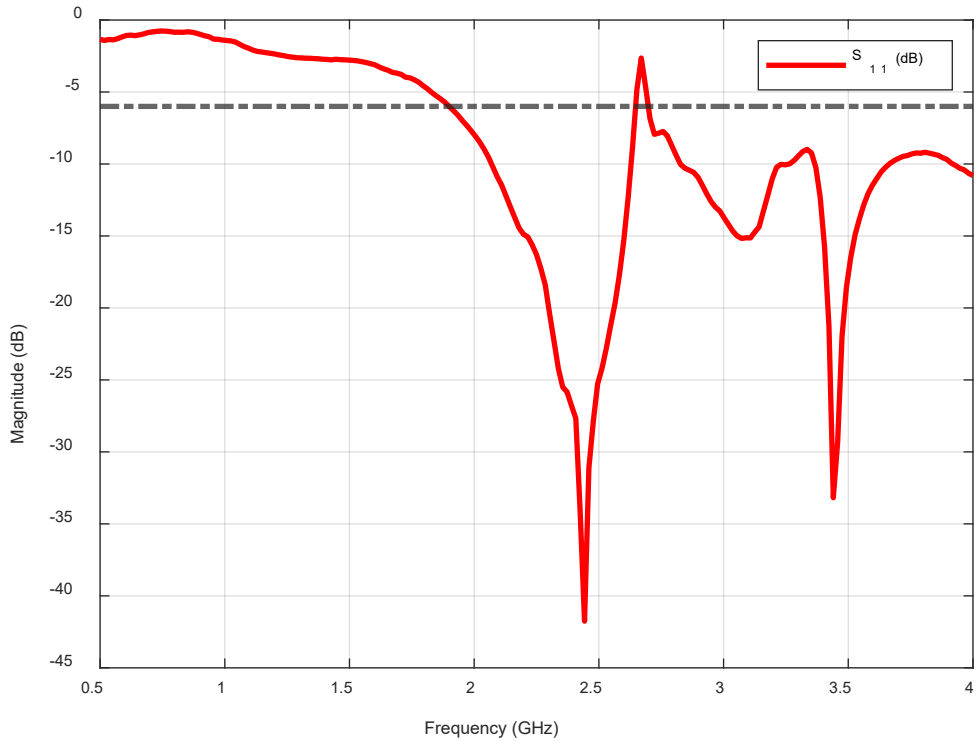

**Figure S2: Return loss ($S_{11}$) of the surface acoustic wave interdigitated transducer used in the experiment for coherent control of NV-defect with nanoscale magnets. The measured data shown in the figure gives a broad bandwidth of transmission of SAW wave including the resonant frequencies of the nitrogen vacancy center under a bias field during the pulsed ODMR measurements. The bias field and position of the NV-center tip is optimized to get a clear ODMR signal in the operating range of the IDT where it is able to transmit with low reflection loss.**

## 2.2 Hysteresis loop of Cobalt films:

To show the magnetic behavior of the nanomagnets that was driven by surface acoustic wave to control the NV-center in diamond, the coercivity of a representative magnetic material is measured, which is used to pattern and deposit the nanomagnets. Titanium (Ti) and cobalt (Co) materials are deposited on top of the lithium niobate substrate following the same method as of the nanomagnets and the hysteresis loop is measured by magneto-optic Kerr microscopy (MOKE) [S1]. The deposited thin film thickness of the Ti and Co are 4 nm and 15 nm, respectively which is similar to the thicknesses of the nanomagnets used in the experiment. The film has no perpendicular magnetic anisotropy (PMA) and is easy to magnetize in the in-plane direction, which saturates at around ~ 3 mT of external magnetic field along the x-direction shown at the bottom right of supplementary figure S3. The film is hard to saturate in the out-of-plane direction and observe high coercivity, showing ferromagnetic characteristics of the Co film.

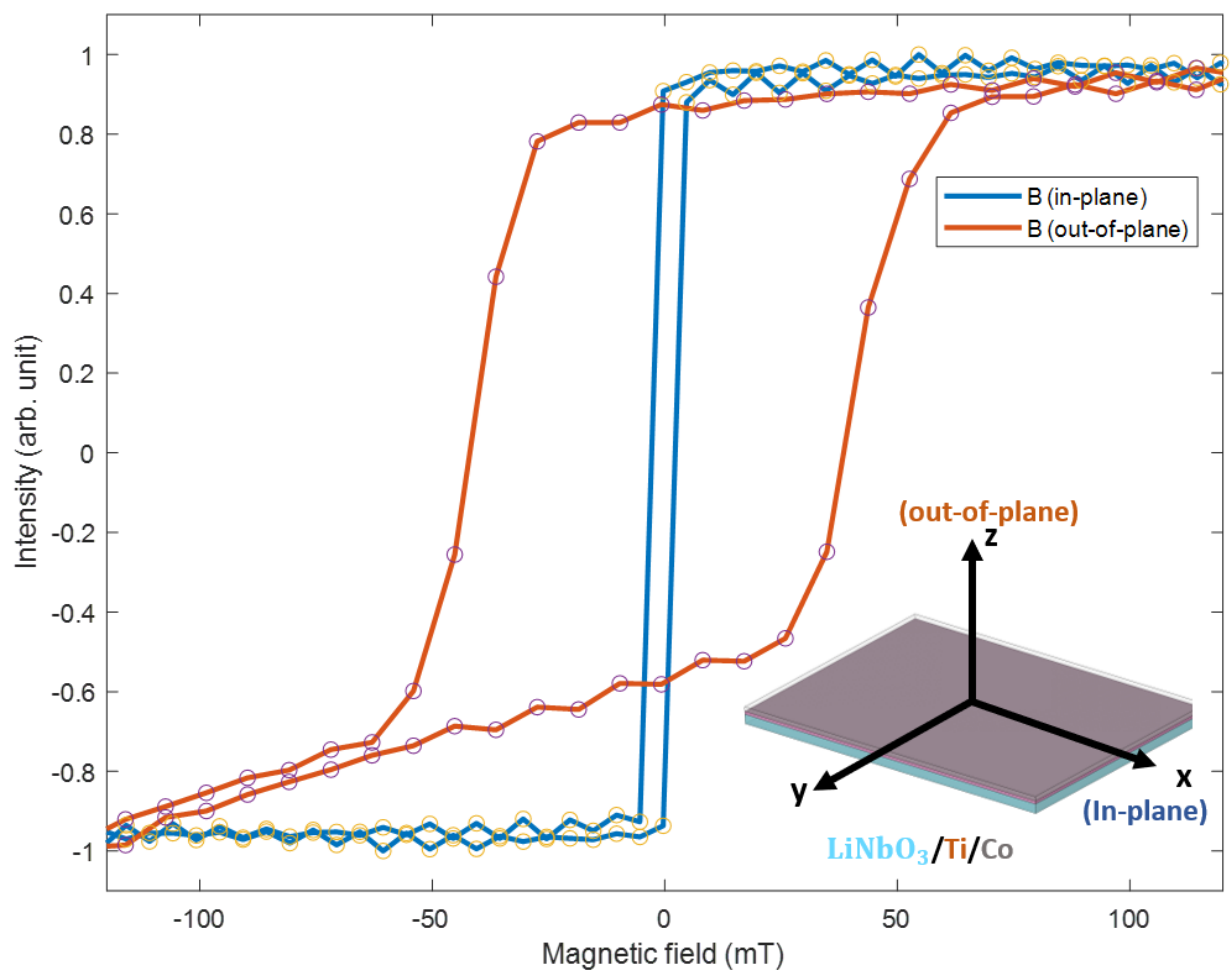

**Figure S3: Magnetic hysteresis loop of 4 nm Ti and 15 nm Co material measured in-plane (IP) and out-of-plane (OP) of the film using MOKE microscopy. The violet scatter plots represent the experimental data for out-of-plance measurement and yellow scatter plot represent in-plane measurement. The schematic (image not in scale) at the bottom right shows the device material stack and direction of the IP and OP magnetic field application during measurement.**

### 2.3 Magnetic Force Microscopy images of Cobalt Nanomagnets:

The nanomagnets are characterized by a tapping magnetic force microscope with a cobalt ferromagnet tip attached to a cantilever with resonant frequency at ~75 kHz. The topography image (height) of the 200 nm x 180 nm is shown in figure S4(a) and the corresponding phase image is shown in figure S4(b). The magnetization is oriented along the easy axis of the nanomagnet as expected. With the increasing of shape anisotropy (i.e. increasing the ratio of the easy axis and hard axis diameter), the magnetization contrast can be further enhanced.

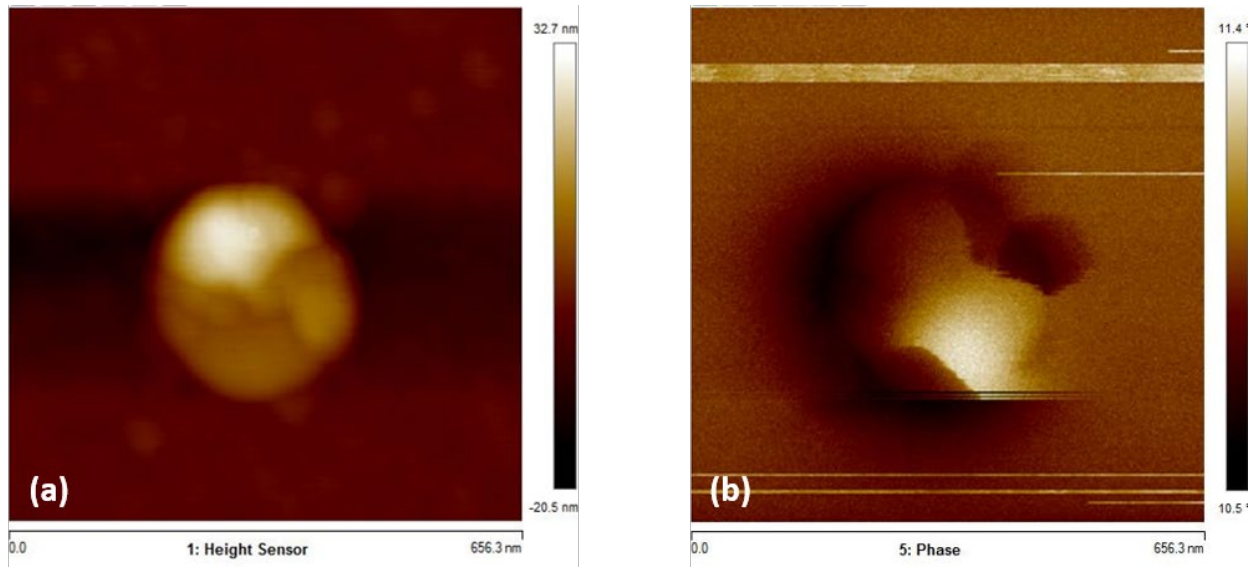

**Figure S4: Magnetic force microscopy (MFM) images of nanomagnets of different sizes. (a) Topography (height) image of a 200 nm x 180 nm nanomagnet (b) Phase image of the corresponding nanomagnet showing magnetic domain orientation along the easy axis as expected.**

The nanomagnets are simulated with MuMax3 micromagnetic analysis software [S2] to see the domain of the nanomagnets in the absence of a magnetic field and also under a bias field. The theory behind the micromagnet simulation is presented in supplementary section in 4.1. Magnetic domains of nanomagnets with different size, thickness and shape anisotropy are shown in supplementary figure S5. The nanomagnets has multiple domains for lower shape anisotropy and can be formed to a single domain nanomagnet with higher shape anisotropy and with application of bias magnetic field in the direction of easy axis of the nanomagnets.

A 50 mT bias field is applied in the same nanomagnets and corresponding domain structure is plotted in the respective box. The magnetization is oriented in the easy axis upon application of a bias field along the same axis of the nanomagnets.

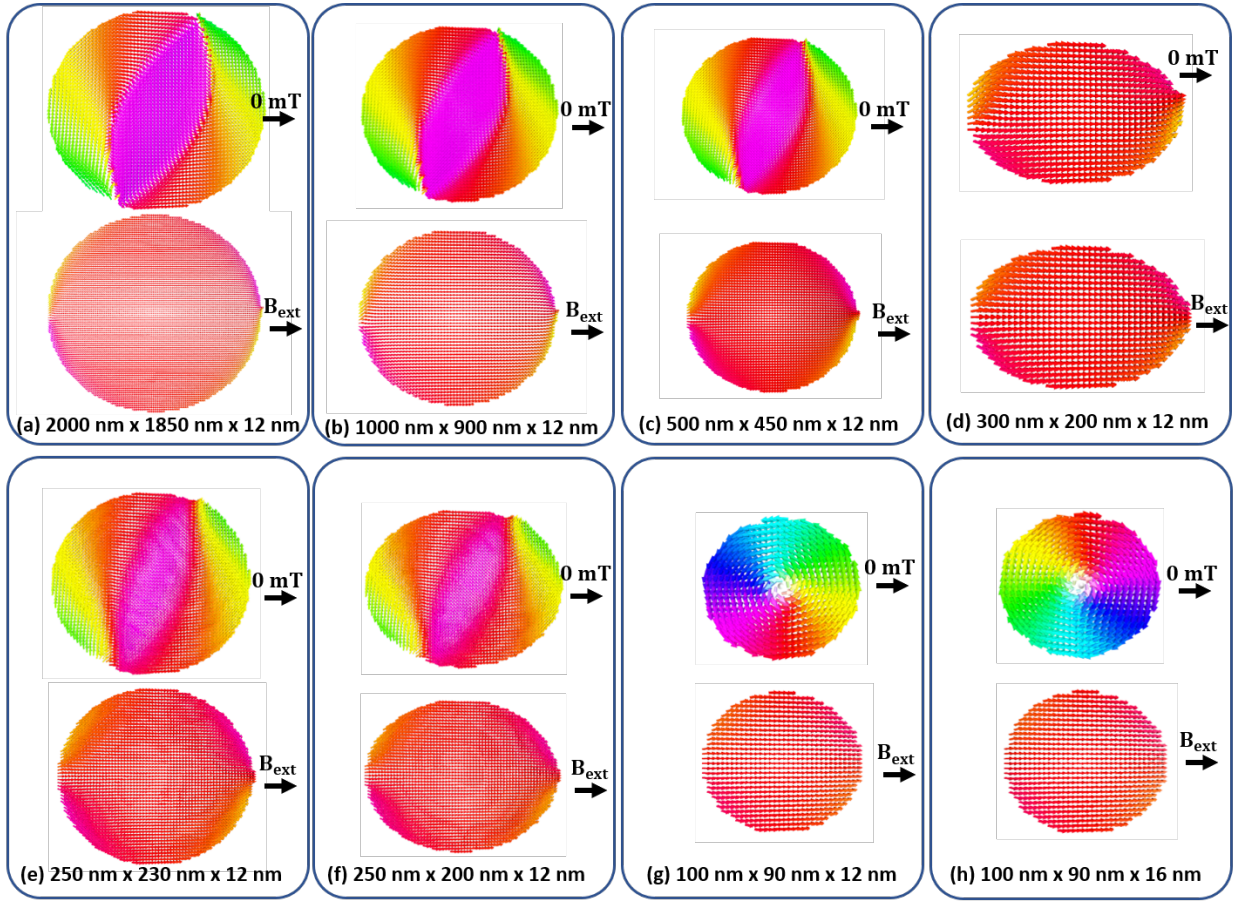

**Figure S5: Magnetic domain analysis of elliptical nanomagnets with different thickness and shape anisotropy. Domain (top view) of (a) 2000 nm x 1850 nm x 12 nm nanomagnet, (b) 1000 nm x 900 nm x 12 nm nanomagnet, (c) 500 nm x 450 nm x 12 nm nanomagnet, (d) 300 nm x 200 nm x 12 nm nanomagnet, (e) 250 nm x 230 nm x 12 nm nanomagnet, (f) 100 nm x 90 nm x 12 nm nanomagnet, (g) 100 nm x 90 nm x 12 nm nanomagnet, and (h) 100 nm x 90 nm x 16 nm nanomagnet. The 50 mT external bias field is applied along the x-axis. (image not in scale)**

### 3. Nitrogen-vacancy center ground state Hamiltonian:

The NV center can be initialized and read-out of the electronic state of the electron can be done optically. Green laser illumination initializes the NV center to $m_s = |0\rangle$ state and fluorescence contrast is measured to estimate the spin state population, which can go up to 30 % of the dark state ($m_s = |0\rangle$). If a microwave field is applied which is resonant to the energy transition between $m_s = |0\rangle$ and $m_s = |\pm 1\rangle$, high contrast is observed in the fluorescence measurement which is basis of optically detected magnetic resonance [S3]. The energy difference can be found from the ground state spin Hamiltonian of the NV-center which is defined as follows:

$$\hat{\mathcal{H}} = \hbar D \hat{S}_{\hat{z}}^2 + \hbar E \left( \hat{S}_{\hat{x}}^2 - \hat{S}_{\hat{y}}^2 \right) + g_e \mu_B B_{NV} \hat{S}_{\hat{z}} + g_e \mu_B B_{\parallel} \hat{S}_{\hat{z}} + g_e \mu_B B_{\perp} . (\hat{S}_{\hat{x}} + \hat{S}_{\hat{y}}) \tag{Se1}$$

where $D$ and $E$ is the NV axis zero-field splitting (ZFS) parameter. The spin-spin interaction in the ground state due to electrons the excited state is formed with ZFS, $D \cong 2.87 \ GHz$. $\hat{S}_{\hat{z}}^2$ represents electron spin interaction with the crystal field along quantization-axis. $E$ is off-axis zero-field splitting parameter and mostly depends on the local strain of the diamond matrix that hosts the NV-center, with a typical value of 5 MHz [S4]. $g_e \simeq 2$ is the electron Lande g-factor, $\mu_B$ is the Bohr magneton calculated by $\mu_B = e\hbar/2m_e$, $\hbar$ is reduced Plank's constant, $B_{NV}$ is the applied magnetic field along the NV-axis, $\hat{S}_{\hat{x}}, \hat{S}_{\hat{y}}, \hat{S}_{\hat{z}}$ are the spin-1 operators.

The NV tip used in the experiment is oriented as shown in supplementary figure S6 with θ=90° and φ=144°. So, the external bias field is applied along the direction of θ=90° and φ=144°. The cobalt nanomagnet has a static field according to it's domain orientation that is applied on the NV-center when the defect becomes closer to the nanomagnet. $B_{\parallel}$ is the parallel component along the NV-axis and $B_{\perp}$ is the longitudinal component of the NV-axis (off-axis component).

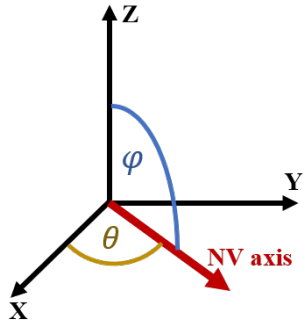

**Figure S6: NV center tip orientation in the diamond used in the coherent control with nanoscale magnet experiment where $\theta = 90°$ and $\varphi = 144°$.**

## 4.    Micromagnet simulations to explore nanomagnet scaling

### 4.1 Micromagnetic simulation:

The simulations of the magnetization dynamics in the nanomagnets are performed by solving the Landau-Lifshitz-Gilbert (LLG) equation using a micromagnetic framework [S2]

$$\frac{d\vec{m}}{dt} = -\frac{1}{(1+\alpha^2)}\gamma[\vec{m} \times \vec{H}_{eff}] - \frac{\alpha}{(1+\alpha^2)}\gamma[\vec{m} \times (\vec{m} \times \vec{H}_{eff})]$$    (Se3)

Here, $\alpha$ is the Gilbert damping coefficient, $\gamma$ is the gyromagnetic ratio, $\vec{m} = \frac{\vec{M}}{M_s}$ is the normalized magnetization, where $\vec{M}$ is the magnetization and $M_s$ is the saturation magnetization.

The effective magnetic field, $\vec{H}_{eff}$ in this case consists of the fields due to the exchange field, uniaxial anisotropy of the nanomagnets, and the demagnetizing field.

$$\vec{H}_{eff} = \vec{H}_{anis} + \vec{H}_{exchange} + \vec{H}_{demag}$$    (Se4)

$\vec{H}_{anis}$ is the effective field due to uniaxial perpendicular magnetic anisotropy (PMA) which can be modulated using Voltage Control of Magnetic Anisotropy (VCMA), $\vec{H}_{exchange}$ is the effective field due to Heisenberg exchange coupling, and $\vec{H}_{demag}$ is the field due to demagnetization energy (shape anisotropy).

$$\vec{H}_{exchange} = \frac{2A_{ex}}{\mu_0 M_s}\sum_i \frac{(\vec{m_k} - \vec{m_c})}{\Delta_i^2}$$    (Se5)

Where $k$ represents the six nearest neighboring cells of the central cell, $\vec{m_c}$ is the magnetization of the central cell

The effective field due to the perpendicular magnetic anisotropy, $\vec{H}_{anis}$ is given as:

$$\vec{H}_{anis} = \frac{2K_{u1}}{\mu_0 M_s}(\vec{z}.\vec{m})\vec{z}.$$    (Se6)

Here, the first order uniaxial anisotropy constant is $K_{u1}$, the magnetic permeability of free space is $\mu_0$, and $\vec{z}$ is the unit vector corresponding to the anisotropy direction.

### 4.2 Downscaling the size of the nano-magnet and stray field proximal to the nanomagnet:

The nanomagnet that drives the NV-center can be downscaled to tens of nm dimension leveraging the state-of-the art nanofabrication technology and can be designed to generate high microwave field for faster Rabi oscillation. The downscaling of the nanomagnet will further enhance the scalability of the quantum system. A micromagnetic simulation of the 10 nm and 100 nm nanomagnets with shape anisotropy is presented in supplementary figure S5. The magnetization dynamics of the nanomagnet is simulated by solving the Landau-Lifshitz-Gilbert (LLG) equation through MuMax3 (see supplementary section 4.1). By applying a sinusoidal perturbation by surface acoustic wave or voltage-controlled-magnetic-anisotropy (VCMA), an oscillating magnetic field is induced in the proximity. The amplitude of the oscillating control field is higher in proximity and reduces with the distance from the nanomagnet, as shown in supplementary figure S7. A schematic of the simulation is shown in figure S7(a), where A and B are two points along x-axis where the magnetic field amplitude is measured. Figure S7(c) and S7(d) show the oscillating field at 5 nm and 10 nm distance from the 10 nm x 7 nm x 2 nm nanomagnet and figure S7(e) and S7(f) shows the oscillating field at 10 nm and 15 nm distance from the 100 nm x 70 nm x 10 nm nanomagnet. For deterministically manipulate the nanomagnet in +x-direction a small magnetic field is applied.

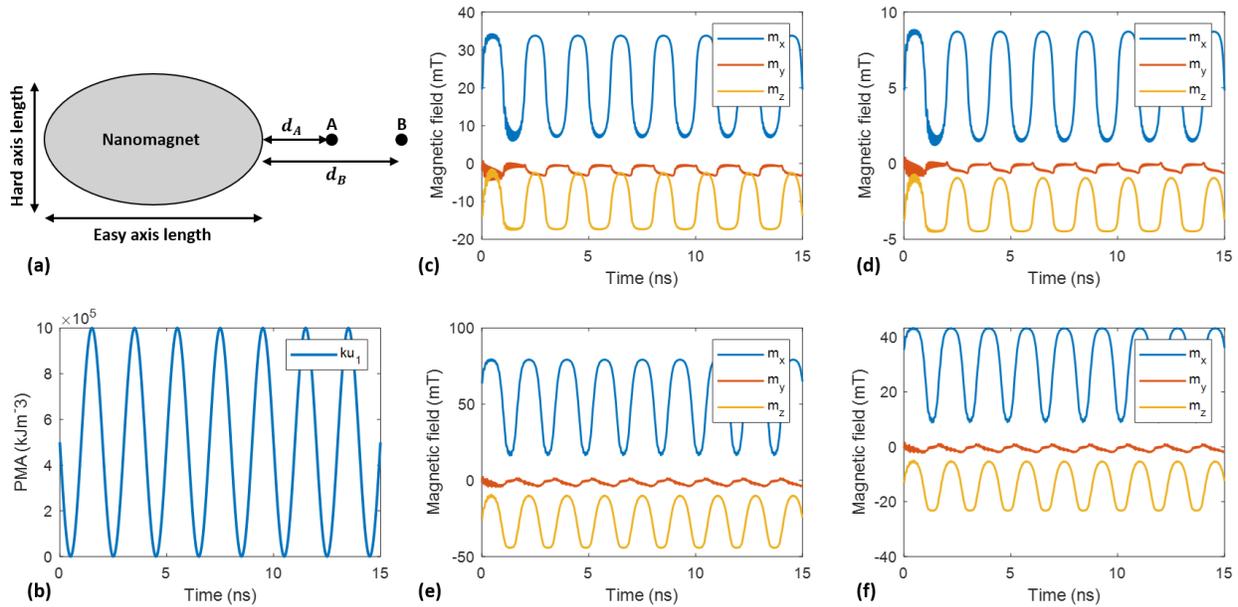

**Figure S7: Downscaling the size of the nano-magnet. (a) A schematic of the simulation in MuMax3. An elliptical nanomagnet is chosen and driven with microwave perturbation at 500 MHz. The stray field is measured at point A and point B, with distances $d_A$ and $d_B$ from the edge of the nanomagnet, respectively. Proximal microwave field from 10 nm x 7 nm x 2 nm and 100 nm x 70 nm x 10 nm nanomagnets at different distances. (b) Applied sinusoidal $ku_1$ to the nanomagnets at 500 MHz frequency. Stray microwave field calculated at (c) 5 nm and (d) 10 nm distance for nanomagnet 10 nm x 7 nm x 2 nm. Stray microwave field calculated at (e) 5 nm and (f) 10 nm distance for nanomagnet 100 nm x 70 nm x 10 nm.**

## 5.  Estimation of energy efficiency of SAW drive vs. electromagnetic antenna (*check and revise

The microwave field generated by a 500 nm x 470 nm nanomagnet with the SAW IDT was able to achieve almost ~9 MHz Rabi oscillation frequency with $\pi/2$ gate duration of ~30 ns and $\pi$ gate duration of ~61 ns. To achieve a similar Rabi frequency of 10 MHz with quantum scanning microscope (QSM) antenna it needs to use multiple orders of magnitude of microwave power. The voltage setpoint to drive the IDT is 0.1 V which corresponds to setting the input voltage amplitude at 0.2 V and an arbitrary waveform generator microwave power factor of 0.5, while the QSM antenna needs a voltage of 2 V. The IDT is multiple orders of energy efficient than that of the optimized QSM antenna, which can be calculated as $-26.02$ dB which corresponds to ~399.9 times gain calculated by equation Se1. Additionally, while the antenna is placed very close (~ 50 µm) to the NV-defect, the surface acoustic IDT fingers are placed ~ 400 µm away from the nanomagnet which is approximately 30 nm from the spin defect.

$$\text{Power gain} = 10 \log\left(\left|\frac{P_{QSM}}{P_{IDT}}\right|\right) = 20 \log\left(\left|\frac{V_{QSM}}{V_{IDT}}\right|\right) \qquad \text{(Se7)}$$

The power dissipation due to SAW excitation on the piezoelectric substrate mostly depends on the potential ($V_s$) applied to induce required stress ($\sigma$) to strain the magnetostrictive layer ($Co$), Young's modulus of Co ($Y$), the IDT beamwidth ($W$), frequency of SAW ($f_{SAW}$), and the piezoelectric substrate ($LiNbO_3$) properties such as $d_{33}$ coefficient (ratio of induced strain to the applied electric field), admittance ($y_a$), SAW propagation speed etc.

The power dissipation by the SAW IDT is defined by [S5]

$$\frac{P}{W} = \frac{1}{2}|V|^2\left(\frac{y_a}{\lambda}\right) \qquad \text{(Se8)}$$

Here $V$ is the required surface potential to induce strain in the nanoscale magnets proximal to the NV center, $W$ is the beamwidth of the IDT, $y_a$ is admittance and $\lambda$ is the wavelength of propagation. For a 2.665 GHz excitation with a 0.1 V applied voltage to the IDT, the calculated propagation wavelength is 1346 nm assuming the SAW propagation speed in $LiNbO_3$ is 3588 m/s. The power dissipation is calculated as 0.7801 W/m. For the IDT beamwidth of 50 µm and for implementing a 100 ns $X$-gate of duration, the energy dissipation per gate is 3.9 x 10^-12 J and the required power is 39 µW, which corresponds to ~-14 dBm. This means that, the IDT and nanomagnet can be optimized to operate at a very low power input compared to existing microwave sources for coherent control of spin systems.

Besides, the power consumption of the SAW IDT used in the experiment can be further minimized with careful optimization of the nanomagnet dimension as well as magnetostrictive materials and optimizing the design parameters of the IDT, thickness of the piezoelectric substrate [S6-S7].

The SAW IDT is used to control the nanomagnet, creating a rotating magnetic field that acts as the control field for NV center qubit gates. Strain-based manipulation of the nanomagnet requires roughly 100 attojoules (aJ) of power to achieve a 90-degree rotation of magnetization [S8-S10]. In contrast, Voltage-Controlled Magnetic Anisotropy (VCMA) requires about 1 femtojoule (fJ) for a perpendicular magnetization switch [S11-S15]. As depicted in the figure, a full 90° rotation is not necessary to produce a microwave field. Instead, we can assume a 20 % (18°) rotation is sufficient to generate a high-amplitude control field that can induce significant Rabi oscillations. Given a qubit π-gate duration of 100 nanoseconds (ns), a 2.5 GHz microwave field requires 250 rotations of the magnetization. Assuming a linear relationship between the power required by the IDT and the degree of magnetization rotation, approximately 2.5 fJ is needed for strain-based manipulation and around 25 fJ for VCMA-assisted manipulation for these operations. Hence, this approach offers multiple orders of magnitude in energy efficiency compared to conventional antennas used to generate microwave fields for controlling qubits.

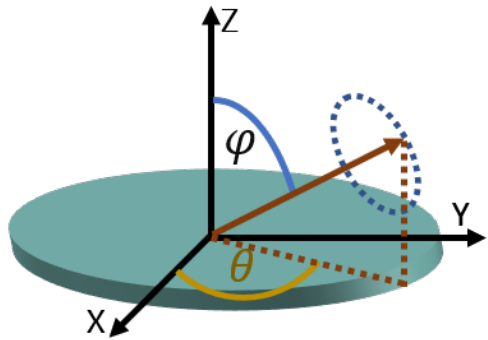

**Figure S8: Magnetization switching of 90° or 180° of the nanomagnets is not necessary to generate microwave control field to the NV center. (image not in scale)**

**Acknowledgement for supplementary materials:** We acknowledge the contribution of Michael Suche and Erdem Topsakal for scattering parameter ($S_{11}$) measurement of SAW IDT with vector network analyzer.

**Supplementary references:**

S1. Shinagawa, K., 2000. Faraday and Kerr effects in ferromagnets. In *Magneto-optics* (pp. 137-177). Berlin, Heidelberg: Springer Berlin Heidelberg.

S2. Vansteenkiste, A., Leliaert, J., Dvornik, M., Helsen, M., Garcia-Sanchez, F. and Van Waeyenberge, B., 2014. The design and verification of MuMax3. *AIP advances*, *4*(10).

S3. Doherty, M.W.; Manson, N.B.; Delaney, P.; Jelezko, F.; Wrachtrup, J.; Hollenberg, L.C. The nitrogen-vacancy colour centre in diamond. Phys. Rep. 2013, 528, 1–45.

S4. Rondin, L., Tetienne, J.P., Hingant, T., Roch, J.F., Maletinsky, P. and Jacques, V., 2014. Magnetometry with nitrogen-vacancy defects in diamond. *Reports on progress in physics*, *77*(5), p.056503.

S5. Datta S (1986) Surface acoustic wave devices. Prentice-Hall, Englewood Cliffs, NJ.

S6. Liu, Yi, and Tianhong Cui. "Power consumption analysis of surface acoustic wave sensor systems using ANSYS and PSPICE." *Microsystem technologies* 13 (2007): 97-101.

S7. Wilson, William C., and Gary M. Atkinson. "Rapid SAW sensor development tools." In *CANEUS/NASA Workshop on Fly-by-Wireless for Aerospace Vehicles*. 2007.

S8. Ahmad, H., Atulasimha, J. & Bandyopadhyay, S. Reversible strain-induced magnetization switching in FeGa nanomagnets: Pathway to a rewritable, non-volatile, non-toggle, straintronic memory cell for extremely low energy operation. Sci. Rep., 5, 18264 (2015).

S9. D'Souza, N., Salehi Fashami, M., Bandyopadhyay, S. & Atulasimha, J. Experimental clocking of nanomagnets with strain for ultralow power Boolean logic. Nano Lett., 16, 1069-1075 (2016).

S10. Zhao, Z. et al. Giant voltage manipulation of MgO-based magnetic tunnel junctions via localized anisotropic strain: Pathway to ultra-energy-efficient memory technology. Appl. Phys. Lett., 109, 092403 (2016).

S11. Wang, K.L., Lee, H. and Amiri, P.K., 2015. Magnetoelectric random access memory-based circuit design by using voltage-controlled magnetic anisotropy in magnetic tunnel junctions. *IEEE Transactions on Nanotechnology*, *14*(6), pp.992-997.

S12. Sengupta, A., Panda, P., Wijesinghe, P., Kim, Y. and Roy, K., 2016. Magnetic tunnel junction mimics stochastic cortical spiking neurons. *Scientific reports*, *6*(1), p.30039.

S13. Grezes, C. et al. Ultra-low switching energy and scaling in electric-field-controlled nanoscale magnetic tunnel junctions with high resistance-area product. Applied Physics Letters. 108, 012403 (2016).

S14. Rajib, M.M., Al Misba, W., Bhattacharya, D., Garcia-Sanchez, F. and Atulasimha, J. Dynamic skyrmion-mediated switching of perpendicular MTJs: feasibility analysis of scaling to 20 nm with thermal noise. IEEE Transactions on Electron Devices. 67, 3883-3888 (2020).

S15. Bhattacharya, D. and Atulasimha, J. Skyrmion-mediated voltage-controlled switching of ferromagnets for reliable and energy-efficient two-terminal memory. ACS Applied Materials & Interfaces. 10, 17455-17462 (2018).